\documentclass[10pt]{article}          %%
\usepackage{mathrsfs,amsfonts}
minus.4em \belowdisplayshortskip=2mm plus.2em minus.4em

\makeatletter 
\@addtoreset{equation}{section}  \makeatother

\title{\centerline{\normalsize} %\hfill hep-th/0603226} 
\bf{Self-adjointness of
generalized  MIC-Kepler system }}

\author{ {\bf Pulak Ranjan Giri\thanks{e-mail :pulakranjan.giri@saha.ac.in}}\\
\normalsize Saha Institute of Nuclear Physics, 1/AF Bidhan-Nagar,  Calcutta
700064, India}

\date{\today}

\begin{document}\maketitle
%%%%%%%%%%%%%%%%%%%%%%%%%
\begin{abstract} \noindent\small
We have studied the self-adjointness of generalized MIC-Kepler Hamiltonian,
obtained  from the  formally self-adjoint generalized MIC-Kepler
Hamiltonian. We have shown that for $\tilde l=0$, the system admits a
1-parameter family of self-adjoint extensions and for $\tilde l \neq 0$ but
$\tilde l <\frac{1}{2}$, it has also a 1-parameter family of self-adjoint
extensions.\\

\noindent {\bf Keywords:} MIC-Kepler system; Self-adjoint extensions
; Bound states\\
\noindent PACS numbers:   03.65.Db, 02.30.Ik, 03.65.-w
\end{abstract}
%%%%%%%%%%%%%%%%%%%%%%%%%
%pacs number 14.80.Hv
%%%%%%%%%%%%%%%%%%%%%%%%%
\section{\small{\bf {Introduction}}} \label{in}
%%%%%%%%%%%%%%%%%%%%%%%%%
Exactly solvable models are extremely important in both theoretical and
mathematical physics because with these model one could get an idea of the the
different physics which could arise in reality.
The non-relativistic 
Coulomb problem is one such example, which  has immense importance 
due to its symmetry under specific transformation. 
It can be shown that the over-complete symmetry gives rise
to the degeneracy in energy spectrum, separability of variables in some
coordinate systems. The Hamiltonian with only Coulomb potential has $O(4)$
symmetry for bound states and $O(1,3)$ symmetry for scattering states. But if
the particles of the Coulomb problem have magnetic charges $g_1$ and $g_2$ in
addition to their electric charges $e_1$ and $e_2$, then usually the  symmetry
is destroyed. It was shown \cite{zwa} that this symmetry can be restored if an
additional scalar potential  of the form $\sim\frac{1}{r^2}$, where $r$ is the
radial distance of the center of mass  from the origin, is added to the
Hamiltonian. This system is known as MIC-Kepler system and is discussed in
Ref. \cite{mcl} also. Apart from $O(4)$ symmetry for bound state and $O(1,3)$
symmetry for continuum state, there are two more features which are preserved
for MIC-Kepler system: (1) the differential cross section remains the same,
both classically and quantum mechanically and (2) if there are three
nonrelativistic particles with both electric and magnetic charges with pair
wise interaction of the form $\sim\frac{1}{r^2}$, then if two particles are
held fixed, the Hamiltonian of the third particle is separable in ellipsoidal
coordinates.

One can obtain MIC-Kepler system from the four dimensional isotropic
oscillator by Kustaanheimo-Stiefel transformation \cite{k1,k2}.  Similarly
from two and eight dimensional isotropic oscillator one can get two
\cite{2d1,2d2} and five \cite{5d1,5d2} dimensional analog of MIC-Kepler
system. It has been generalized from different perspective. For example,  we
can have generalization of MIC-Kepler system on a class of three dimensional
spaces having conic singularities from quantum mechanical oscillator defined
on K\"{a}hler conifold \cite{bellucci}. There is another generalization
\cite{mar},  by adding Smorodinsky-Winternitz type potential 
\cite{fris1,fris2,fris3} to
the kinetic energy term of the Hamiltonian. The Hamiltonian $H(s, c_1, c_2)$
\cite{mar} dependents on the parameters $s, c_1, c_2$. The
Smorodinsky-Winternitz type potentials where studied in Ref. 
\cite{evan1,evan2,evan3}. This
potential can be reduced to Hartmann potential \cite{hart1,hart2,hart3}, which has been
used for describing axially symmetric systems like ring-shaped molecules and
investigated from different point of view 
\cite{kib1,kib2,kib3,kib4,kib5,kib6,kib7,kib8,kib9,kib10,kib11,kib12,kib13}. 
This system has rich
features, i.e., exact solvability, separability of variables in few coordinate
systems.
It is solved in various coordinate systems \cite{mar2}, for example;
spherical, parabolic and prolate spheroidal coordinate systems.  Interesting
feature of this generalized system $( H(s, c_1, c_2))$ is that it can be
reduced to many interesting known quantum mechanical systems. For $c_1=c_2=0$,
it reduces to MIC-Kepler system \cite{zwa,mcl}.  For $s=c_1=c_2 = 0 $ in
generalized MIC-Kepler system we  get the Hydrogen atom Hamiltonian
\cite{park,tar,aruty}. Generalized Kepler-Coulomb system \cite{kibler} can be
found  by making $s=0,c_i\ne0~ (i=1,2)$ in generalized MIC-Kepler Hamiltonian
$H(s, c_1, c_2)$ \cite{mar}.  We can get Hartmann system \cite{lut}, when
$s=0,c_1=c_2\ne0$ and also charge-dyon system \cite{mar3} when
$s\ne0,c_1=c_2=0$.

The purpose of this paper is to study self-adjointness of generalized
MIC-Kepler system.  Self-adjointness of similar type problems  have been
studied in many papers. For example, it is shown in Ref.~\cite{few,rel} that the
radial Hamiltonian $h_l$ of the Hydrogen atom with effective potential
$V_{eff}= \frac{\gamma}{r} + \frac{l(l+1)}{r^2}$ has one parameter family of
self-adjoint extensions only for angular momentum quantum number $l=0$ 
and for all
other values of $l$, since deficiency indices are $<0,0>$, it is essentially
self-adjoint. Actually if one demands the square integrability of the
deficiency space solution $u^\pm_l$ at origin, one should get the condition $-
\frac{3}{2}< l < \frac{1}{2}$. Outside this limit since the solutions $u^\pm_l$ are
not square integrable, the   deficiency indices are $<0,0>$. So Hamiltonian is
essentially self-adjoint. Since for Hydrogen atom problem $l$ can have only
nonnegative integral values, it is sufficient to say that for $l\geq 1$ the
radial Hamiltonian is essentially self-adjoint. In \cite{callias} it has been
shown that the Dirac Hamiltonian of an isospin-1/2 particle in the field of
$SU(2)$ monopole  admits a one-parameter family of self-adjoint extensions.  In
case of Calogero model  effective potential of the form $\frac{\tilde{g}}{r^2}$
appears when written in terms of radial variable $r$. Self-adjointness of this
model  has been studied in Ref. \cite{biru1,biru2} and it is shown that for
$-\frac{1}{4} \leq \tilde{g}< \frac{3}{4}$, the Hamiltonian has self-adjoint
extensions. Outside this domain since the Hamiltonian is essentially
self-adjoint, there is no scope for self-adjoint extensions. In our case, the
generalized MIC-Kepler system has effective potential $V_{eff}= -\frac{1}{r} +
\frac{\tilde l(\tilde l +1)}{r^2}$. The difference in radial Hamiltonian
between generalized MIC-Kepler system and Hydrogen atom problem is in the
coefficient of the potential $\sim \frac{1}{r^2}$. In case of Hydrogen atom
$l$ can take values $0, 1,2,3,\cdots$, but in case of generalized MIC-Kepler
system, $\tilde l$ can take other positive values also, besides
$0,1,2,3,\cdots$ (in fact $\tilde l$ can have positive continuous values). It
will allow us to get self-adjoint extensions of the radial Hamiltonian for
$\tilde l\neq 0$ besides $\tilde l=0$, as opposed to Hydrogen atom problem,
where only $l= 0$ has self-adjoint extensions. This is not strange, as we can
see for Calogero model we get self-adjoint extensions for $\tilde g \neq 0$
\cite{biru1,biru2}.

The paper is organized as follows: In Sec. (\ref{so}), we discuss the
generalized  MIC-Kepler system \cite{mar} briefly. In Sec. (\ref{se}), we
study  the self-adjointness issue of this  system. We conclude in
Sec. (\ref{con}).
%%%%%%%%%%%%%%%%%%%%%%%%%%%%%%%%%%%%%%%%%%%%%%%%%%%%%%%%

\section{\small{\bf{Generalized MIC-Kepler system}}}\label{so}
The generalized MIC-Kepler Hamiltonian \cite{mar}, in $\hbar = c = e = 1$
 unit and  with mass taken to be unity is given by,
\begin{eqnarray}
H(s, c_1, c_2) = \frac{1}{2}\left( -i {\bf\nabla} - s {\bf A}\right)^2
+\frac{s^2}{2r^2} - \frac{1}{r} +\frac{c_1}{r^2(1+\cos{\theta})}
+\frac{c_2}{r^2(1 -\cos{\theta})}.
\label{gmicKepler}
\end{eqnarray}
Here ${\bf A}$ is the magnetic vector potential  of the Dirac monopole, given
by
\begin{eqnarray}
{\bf A}= -\frac{\sin\theta}{r(1-\cos\theta)}\hat\phi, \label{pot}
\end{eqnarray}
such that $curl{\bf A} = \frac{\bf r}{r^3}$. $c_1, c_2$ are nonnegative
constants and $s$ takes the values $0, \pm \frac{1}{2},\pm 1, \cdots$.

We now want to study the eigenvalue problem of the Hamiltonian Eq.
(\ref{gmicKepler}), given by
\begin{eqnarray}
H\Psi = E\Psi.
\label{eigen}
\end{eqnarray}
This equation can be easily separated out in radial and angular part in
spherical polar coordinate system, if we consider the wave function $\Psi$ of
the form
\begin{eqnarray}
\Psi(r,\theta, \phi) = R(r)\Phi(\theta, \phi).
\label{eigenfun}
\end{eqnarray}
Substituting Eq. (\ref{eigenfun}) into  Eq. (\ref{eigen}) we get two decoupled
differential equations. The differential equation involving angular
co-ordinates is of the form
\begin{eqnarray}
\frac{1}{\sin\theta}\frac{\partial}{\partial\theta}\left(\sin\theta\frac{\partial
\Phi}{\partial\theta}\right) +
\frac{1}{4\cos^2\frac{\theta}{2}}\left(\frac{\partial^{2}}{\partial\varphi^{2}}-4c_1\right)\Phi+
\nonumber\\
\frac{1}{4\sin^2\frac{\theta}{2}}\left[\left(\frac{\partial}{\partial\varphi}+2is\right)^{2}
-4c_2\right]\Phi = -{\tilde l(\tilde l +1)}\Phi, \label{angular}
\end{eqnarray}
where $\tilde l(\tilde l+1)$ is the separation constant. Solution of
Eq. (\ref{angular}) is well discussed in Ref.~\cite{mar}, and is of the form
\begin{eqnarray}
\Phi_{jm}^{(s)}(\theta, \varphi; \delta_{1}, \delta_{2} )=N_{jm}(\delta_{1},
\delta_{2})\left(\cos\frac{\theta}{2}\right)^{m_1}
\left(\sin\frac{\theta}{2}\right)^{m_2} P_{j-m_+}^{(m_2,m_1)}(\cos\theta)
e^{i(m-s)\varphi}.
\label{angularsol}
\end{eqnarray}
Here
\begin{eqnarray}
m_1=|m-s|+\delta_{1}=\sqrt{(m-s)^2+4c_1}\nonumber,\\
m_2=|m+s|+\delta_{2}=\sqrt{(m+s)^2+4c_2},
\label{m}
\end{eqnarray}
and
\begin{eqnarray}
m_+=(|m+s|+|m-s|)/2.
\label{m_+}
\end{eqnarray}
$P_n^{(a,b)}$ is a Jacobi polynomial. For a fixed value of $j$ the quantum
number $m$  runs through values:
\begin{eqnarray}
m=-j,-j+1,\dots,j-1,j,
\label{mvalue}
\end{eqnarray}
 whereas the allowed values of $j$ are
\begin{eqnarray}
j = |s|,  |s|+1, |s|+2,  \dots.
\label{jvalue}
\end{eqnarray}
The separation constant $\tilde l$ is of the form
\begin{eqnarray}
\tilde{l}= j+\frac{\delta_{1}+\delta_{2}}{2}.\label{angquantum}
\end{eqnarray}
From Eq. (\ref{m}) and Eq. (\ref{jvalue}) it is clear that $\tilde l$ can have
only positive values. $N_{jm}(\delta_{1},\delta_{2})$ is the normalization
constant. The radial part of the differential equation obtained from
Eq. (\ref{eigen}),  Eq. (\ref{eigenfun}) and  Eq. (\ref{angular}) is of the
form
\begin{eqnarray}
H_{\tilde l}(r)R(r) = ER(r), \label{radialeigen}
\end{eqnarray}
where the effective formal radial Hamiltonian  $H_{\tilde l}(r)$ is
\begin{eqnarray}
H_{\tilde l}(r) = -\frac{1}{2}\frac{1}{r^{2}}
\frac{d}{dr}\left(r^{2}\frac{d}{dr}\right) - \frac{1}{r} +\frac{\tilde
l(\tilde l+1)}{2r^{2}}. \label{radialhamil}
\end{eqnarray}
The bound state solution of the eigenvalue problem Eq. (\ref{eigen}) is
previously known \cite{mar}. However since the Hamiltonian self-adjointness of
Eq. (\ref{radialhamil}) is  important and the operator Eq. (\ref{radialhamil})
is not self-adjoint in a chosen domain,  we will solve the eigenvalue problem
by making  a self-adjoint extensions of the radial Hamiltonian in the next
section.

The generalized MIC-Kepler system can easily be reduced to MIC-Kepler system
By making  $c_1=c_2=0$ and  MIC-Kepler  system is known to be   $O(4)$
invariant for the bound state. It  should therefore  be relevant to  mention
the generators    of the $O(4)$  group  for MIC-Kepler system here. The
angular momentum and the Runge-Lenz vector are given respectively by
\begin{eqnarray}
\hat{{\bf J}}& =&{\bf r}\times\left(-i\nabla- s{\bf A}\right)- s\frac{\bf
r}{r}\\ \hat{{\bf I}}& =&\frac{1}{2}\left[\left(-i\nabla- s{\bf
A}\right)\times{\bf J} - {\bf J}\times \left(-i\nabla- s{\bf A}\right)
\right]+ \frac{\bf r}{r}.
\label{generator}
\end{eqnarray}
The generators $\hat{{\bf J}}$, $\hat{{\bf I}}$ together with the Hamiltonian
$H(s,c_1,c_2)$ form the $o(4)$ algebra. In case of generalized MIC-Kepler
system since two extra potentials with coefficients $c_1$ and $c_2$ are added
with the MIC-Kepler Hamiltonian, the constants of motion should be
modified. The square of the generalized angular momentum is given by
\begin{eqnarray}
{\bf \mathcal J}^2= \hat{{\bf J}}^2 + \frac{2c_1}{1+\cos\theta}+
\frac{2c_2}{1-\cos\theta},
\label{sqangular}
\end{eqnarray}
with eigenvalue
$\left(j+\frac{\delta_1+\delta_2}{2}\right)\left(j+\frac{\delta_1+\delta_2}{2}+1\right)$.
The $z$ component of the angular momentum remains the same
\begin{eqnarray}
\hat{J_z}=s- i\frac{\partial}{\partial\phi}.
\end{eqnarray}
But the Runge-Lenz vector gets modified. The $z$ component of the modified
Runge-Lenz vector is given by
\begin{eqnarray}
{\bf \mathcal I_z}=  I_z + c_1\frac{r-z}{r(r+z)}- c_2\frac{r+z}{r(r-z)}.
\label{runge}
\end{eqnarray}
%
%%%%%%%%%%%%%%%%%%%%%%%%%

%%%%%%%%%%%%%%%%%%%%%%%%%
\section{\small{\bf{Study of self-adjointness of radial differential Hamiltonian}}}\label{se}
We now move to the analysis of the effective differential operator
Eq. (\ref{radialhamil}). For simplicity of calculation we remove the first
order derivative term in Eq. (\ref{radialhamil}). It can be done by the
transformation $R(r)\to \frac{\chi(r)}{r}$ in Eq. (\ref{radialeigen}) and the
resulting radial Hamiltonian becomes
\begin{eqnarray}
\mathcal H_{\tilde l}(r) = -\frac{1}{2}\frac{d^2}{dr^2} - \frac{1}{r}
+\frac{\tilde l(\tilde l+1)}{2r^{2}}\,,
\label{rham}
\end{eqnarray}
which is formally self-adjoint \cite{dunford}.

To study the self-adjointness of this operator we first need to associate this
formal differential operator $\mathcal H_{\tilde l}(r)$ to a closed symmetric
operator $T\left\{\mathcal H_{\tilde l}(r)\right\}$ whose domain $D(T)$ has to
be chosen suitably.  The operator $T\left\{\mathcal H_{\tilde l}(r)\right\}$
in the symmetric domain $D(T)$, takes the same form of Eq. (\ref{rham}).
To find out the domain  $D(T)$ we need to consider the symmetry condition
\begin{eqnarray}
\int_{0}^{\infty}\phi_1(r)^\dagger T\left\{\mathcal H_{\tilde
l}(r)\right\}\phi_2(r)dr = \int_{0}^{\infty} \left[T\left\{\mathcal H_{\tilde
l}(r)\right\}\phi_1(r)\right]^\dagger \phi_2(r)dr\,,
\label{scondition1}
\end{eqnarray}
which can be satisfied if the functions $\phi_1(r),\phi_2(r)$ satisfy the
condition
\begin{eqnarray}
\left(\phi_1(r)\phi'_2(r)\right)_{r=0} -
\left(\phi'_1(r)\phi_2(r)\right)_{r=0}=0\,.
\label{scondition2}
\end{eqnarray}
This can be easily satisfied if $\phi_1(r),\phi_2(r)$ belong to the   domain
\begin{eqnarray}
D(T) = \{\phi(r): \parbox[t]{9cm}{\mbox{$\phi(r=0) = \phi'(r=0) = 0 $},
  absolutely continuous, square integrable  on the  half  line  with  measure
  \mbox{$dr$} \}\,.}
\label{domain1}
\end{eqnarray}
Here $\phi'(r)$ is the derivative of  $\phi(r)$ with respect to $r$.
In order to ensure that the function $T\left\{\mathcal H_{\tilde
l}(r)\right\}\phi(r)$ is square integrable at $r=0$, we need
$\phi''(r)$(second order derivative of $\phi(r)$) to be square
integrable at $r=0$. Which means near the origin  the function
should behave as  $\phi(r)\sim r^k$, where $k> \frac{3}{2}$.  Note
that square integrability of $\phi''(r)$ at $r=0$ automatically
ensure the condition $\phi(r=0) = \phi'(r=0) = 0 $, which has been
used in the above Eq. (\ref{domain1}). We can easily check that
$T\left\{\mathcal H_{\tilde l}(r)\right\}\phi(r)$ is indeed square
integrable at $r=0$ from
\begin{eqnarray}
\left(T\left\{\mathcal H_{\tilde l}(r)\right\}\phi(r)\right)^\dagger
T\left\{\mathcal H_{\tilde l}(r)\right\}\phi(r) dr\sim \nonumber\\
\left[\left(\phi''(r)\right)^2 + \frac{\left(\phi(r)\right)^2}{r^2}+
\frac{\left(\phi(r)\right)^2}{r^4}+
\frac{\left(\phi(r)\right)^2}{r^3}+
\frac{\phi(r)\phi''(r)}{r}+\frac{\phi(r)\phi''(r)}{r^2}\right]dr,
\label{}
\end{eqnarray}
at $r=0$, using the above mentioned condition $\phi(r)\sim r^k$, where $k>
\frac{3}{2}$.

To study the self-adjointness of $T\{\mathcal H_{\tilde l}(r)\}$
with domain $D(T)$ \cite{reed} we have to look for the solution of
the equation
\begin{eqnarray}
T^\dagger\{\mathcal H_{\tilde l}(r)\}\phi^\pm = \pm i\phi^\pm\,,
\label{imaginarysol}
\end{eqnarray}
where $T^\dagger\{\mathcal H_{\tilde l}(r)\}$ is the adjoint of $T\{\mathcal
H_{\tilde l}(r)\}$. The differential form of the adjoint operator
$T^\dagger\{\mathcal H_{\tilde l}(r)\}$ remains the same as
Eq. (\ref{rham}). This can be easily checked from the Green's formula
\cite{dunford}, which relate operator with its adjoint. The domain of the
adjoint operator $T^\dagger\{\mathcal H_{\tilde l}(r)\}$ can be calculated from
\begin{eqnarray}
\int_{0}^{\infty}\phi(r)^\dagger T\left\{\mathcal H_{\tilde
l}(r)\right\}\zeta(r)dr = \int_{0}^{\infty} \left[T^\dagger\left\{\mathcal
H_{\tilde l}(r)\right\}\phi(r)\right]^\dagger \zeta(r)dr\,~~~~~\forall
\zeta\in D(T)\,.
\label{scon1}
\end{eqnarray}
The relation Eq. (\ref{scon1}) will be satisfied if $\phi(r)$ ( $\phi(r)$
should belong to the adjoint domain $D(T^\dagger)$, which we need to find out)
and $\zeta(r)\in D(T)$ satisfy the condition
\begin{eqnarray}
\left(\phi(r)\zeta'(r)\right)_{r=0} - \left(\phi'(r)\zeta(r)\right)_{r=0}=0\,.
\label{scon2}
\end{eqnarray}
From Eq. (\ref{domain1}) and Eq. (\ref{scon2}) we can see that  $\phi(r)$
should belong to the most general domain
\begin{eqnarray}
D(T^\dagger) = \{\phi(r): \parbox[t]{9cm}{ absolutely continuous, square
    integrable  on the  half  line  with  measure \mbox{$dr$} \}\,,}
\label{domain2}
\end{eqnarray}
in order to satisfy Eq. (\ref{scon1}). Obviously $T\left\{\mathcal H_{\tilde
l}(r)\right\}$ is not self adjoint, because
\begin{eqnarray}
D(T)\ne D(T^\dagger)\,.
\label{nonselfad}
\end{eqnarray}
Now let us return to the Eq. (\ref{imaginarysol}).  There are two
solutions
%(these solutions can be found from Eq. (\ref{radialeigen}) by
%replacing $E= \pm i$ and multiplying the resulting eigenfunction by
%$2\varepsilon^\pm r$)
for each differential equation Eq. (\ref{imaginarysol}). But taking square
integrability (at infinity) into account we have to discard one solution. The
remaining one,  apart from normalization, is found to be
\begin{eqnarray}
\phi^\pm(2\varepsilon^\pm r) = (2\varepsilon^\pm
r)^{\frac{b}{2}}e^{-\varepsilon^\pm r}U\left(a^\pm; b; 2\varepsilon^\pm
r\right),
\label{phipmsol}
\end{eqnarray}
where $\varepsilon^\pm = \sqrt{\mp2 i}$ (we should take  $\varepsilon^\pm =
\sqrt{2}e^{\mp i\pi/4}$ here, so that $\phi^\pm(2\varepsilon^\pm r)$ of
Eq. (\ref{phipmsol}) does not blow up at spatial infinity), $a^\pm =j+1
+\frac{\delta_1+\delta_2}{2} -\frac{1}{\varepsilon^\pm}$ and $b=2j+\delta_1+
\delta_2+2 $.  $U$ is the Kummer's  second function  \cite{abro} and its form
is  given  as:
\begin{eqnarray}
U\left(a, b, z\right) = \frac{\pi}{\sin(\pi b)}\left[\frac{ M (a, b,
    z)}{\Gamma(1+a-b)\Gamma(b)} -z^{1-b}\frac{ M (1+a-b, 2- b,
    z)}{\Gamma(a)\Gamma(2-b)} \right]\,,
\label{kummerdef}
\end{eqnarray}
where $M (a, b, z)$ is the  Kummer's first function \cite{abro}.

To check square integrability of Eq. (\ref{phipmsol}) at infinity we have to
consider asymptotic value \cite{abro} of the function $U(2\varepsilon^\pm r)$,
which is in the limit $ r\to\infty$ given by
\begin{eqnarray}
U(a,b,2\varepsilon^\pm r ) \to (2\varepsilon^\pm r)^{-a}\left[ 1+
  \mathcal{O}\{(2\varepsilon^\pm r)^{-1}\}\right]\, \mbox{for} ~~~~\mathcal
  Re(2\varepsilon^\pm r)>0\,.
\label{assymvalue}
\end{eqnarray}
So, solutions of Eq. (\ref{phipmsol}) go to zero as $r$ goes to infinity due to
 the decaying exponential terms  $e^{-\varepsilon^\pm r}$.

Now in the limit \cite{abro} $ r\to 0$, $M (a, b, z)\to 1$. Together with
Eq. (\ref{phipmsol}) and Eq. (\ref{kummerdef}), this implies that in the limit
$r\to 0$,
\begin{eqnarray}
|\phi^\pm(2\varepsilon^\pm r)|^2 dr\to \left[ C_1 r^{2-b} + C_2 r + C_1r^{b}+
 h.o \right]dr\,,
\label{limitzero}
\end{eqnarray}
where $C_1, C_2, C_3$ are constants and $h.o$ is higher order (in powers of
$r$)  terms. Note that the nontrivial coefficient $C_1$ is nonzero here.  From
Eq. (\ref{limitzero}) it is clear that we can get square integrable  solutions
Eq.  (\ref{phipmsol}) only  for $-1 < b < 3$, which implies:
\begin{eqnarray}
-\frac{3}{2}<\tilde l=j+ \frac{\delta_1+\delta_2}{2}<+\frac{1}{2}\,.
\label{squareint}
\end{eqnarray}
But since $\tilde l$ can take only positive values, the effective range for
square integrability becomes
\begin{eqnarray}
0\leq\tilde l=j+ \frac{\delta_1+\delta_2}{2}<+\frac{1}{2}\,,
\label{msquareint}
\end{eqnarray}
which is what we get in case of Hydrogen atom problem. The only difference is
that, in case of Hydrogen atom problem $\tilde l$ in Eq. (\ref{msquareint})
should be replaced by angular momentum quantum number $l$, which can take non
negative integral values only.  Obviously for Hydrogen atom problem only $l=0$
states belong to the specified range Eq. (\ref{msquareint}) and we can perform
self-adjoint extensions for $l=0$ states only.

Beyond the range Eq. (\ref{msquareint}) there is no square integrable solution
of Eq. (\ref{imaginarysol}).  The existence of these complex eigenvalues in
the range Eq. (\ref{msquareint}) of $T^\dagger$ signifies that $T$ is not
self-adjoint in that range.  The solution $\phi^\pm$ belong to the null space
$D^\pm$ of $T^\dagger \mp i$, where $D^\pm \subset D(T^\dagger)$. The
dimension of $D^\pm$ are known as deficiency indices $n^\pm$ and is defined by
\begin{eqnarray}
n^\pm =  dim(D^\pm)\,.
\label{deficiencyindices}
\end{eqnarray}
Depending upon $n^\pm$,  $T$ is classified as \cite{reed}:\\ 1) $T$ is
 self-adjoint if $n^+= n^- = 0$.\\ 2) $T$ has a $n^2$-parameter(real) family
 of self-adjoint extensions if $n^+ = n^-= n \ne 0$.\\ 3) $T$ has no
 self-adjoint extensions if $n^+\ne n^-$.

Since in the range  Eq. (\ref{msquareint}), the deficiency indices $n^+ = n^-
= 1$, which is evident from Eq. (\ref{phipmsol}), we can have a 1-parameter
family of self-adjoint extensions of $T$. The self-adjoint   extensions  of
$T$ is given by $T^\omega$ with domain $D(T^\omega)$, where
\begin{eqnarray}
D(T^\omega)= \{ \psi(r)= \phi(r)+ \phi^+(r) + e^{i\omega}\phi^-(r) :
    \phi(r)\in D(T), \omega\in \mathbb{R} (\bmod 2\pi)\}\,.
\label{selfdomain}
\end{eqnarray}
Beyond the range  Eq. (\ref{msquareint}) the operator $T$ is self-adjoint.
The analysis of the spectrum is same as discussed in many papers
\cite{mar,mar2,pogo}.

%%%%%%%%%%%%%%%%%%%%%%%%%

%%%%%%%%%%%%%%%%%%%%%%%%%
%\section{\small{\bf{Bound state of The Effective Hamiltonian $T^\omega$}}}
%\label{bound}
The self-adjoint radial Hamiltonian $T^\omega\{\mathcal H(r)\}$ is given by
the right hand side of Eq. (\ref{rham}) and the domain is given by
Eq. (\ref{selfdomain}).  Bound state solution of this operator $T^\omega$ is
given by, apart from normalization,
\begin{eqnarray}
\chi(r) = (2\varepsilon r)^{\frac{b}{2}}e^{-\varepsilon r}U\left(a; b;
2\varepsilon r\right),
\label{boundstate}
\end{eqnarray}
where $\varepsilon = \sqrt{-2E}$, $a =j+1 +\frac{\delta_1+\delta_2}{2}
-\frac{1}{\varepsilon} $ and $b=2j+\delta_1+ \delta_2+2 $. To find out the
eigenvalue we have to match the function $\chi(r)$ with the domain
Eq. (\ref{selfdomain}) at $r\to 0$. In the limit $r\to 0$,
\begin{eqnarray}
\chi(r) \to C \left[-\frac{(2\varepsilon)^{-\tilde l}
  }{\Gamma(a)\Gamma(2-b)}(r)^{-\tilde l} + \frac{(2\varepsilon)^{1-\tilde
  l}}{\Gamma(a)\Gamma(2-b)}\left(\varepsilon
  -\frac{1+a-b}{2-b}\right)(r)^{1-\tilde l} \right.\nonumber\\ \left. +
  \frac{(2\varepsilon )^{\tilde l +1}}{\Gamma(1+a-b)\Gamma(b)}(r)^{\tilde l
  +1} + \mathcal{O}(r^{2-\tilde l})\right]\,,
\label{matching1}
\end{eqnarray}
and
\begin{eqnarray}
\phi(r) + \phi^+(r) + e^{i\omega}\phi^-(r) \to D \left[
-\frac{1}{\Gamma(2-b)}\left(\frac{(2\varepsilon^+)^{-\tilde l}}{\Gamma(a^+)}
+\frac{e^{i\omega}(2\varepsilon^-)^{-\tilde l}}{\Gamma(a^-)}
\right)(r)^{-\tilde l} \right.\nonumber\\ \left.
+\frac{1}{\Gamma(2-b)}\left(\frac{(2\varepsilon^+)^{1-\tilde l}}{\Gamma(a^+)}
\left(\varepsilon^+ - \frac{1+ a^+ -b}{2 -b}\right)
+\frac{e^{i\omega}(2\varepsilon^-)^{1-\tilde l}}{\Gamma(a^-)}
\left(\varepsilon^- - \frac{1+ a^- -b}{2 -b}\right) \right)(r)^{1-\tilde
l}\right.\nonumber\\ \left.  +
\frac{1}{\Gamma(b)}\left(\frac{(2\varepsilon^+)^{\tilde l +1}}{\Gamma(1+a^+
-b)} +\frac{e^{i\omega}(2\varepsilon^-)^{\tilde l +1}}{\Gamma(1+a^-
-b)}\right)(r)^{\tilde l +1} + \mathcal{O}(r^{2-\tilde l}) \right]\,.
\label{matching2}
\end{eqnarray}
It should be noted that, in Eq. (\ref{matching2}) there is no contribution
from $\phi(r)$, because if we make Taylor series expansion of  the function
$\phi(r)$ near $r=0$ and use the definition of the domain Eq. (\ref{domain1})
we get the the leading order term  $\phi(r)= \mathcal{O} (r^2)$, which has
been ignored in the right hand side in Eq. (\ref{matching2}).  Equating
coefficients of same powers of $r$ in Eq. (\ref{matching1}) and
Eq. (\ref{matching2}) we get
\begin{eqnarray}
C\frac{(2\varepsilon)^{-\tilde l} }{\Gamma(a)}& = & D\left
  [\frac{(2\varepsilon^+)^{-\tilde l}}{\Gamma(a^+)}
  +\frac{e^{i\omega}(2\varepsilon^-)^{-\tilde l}}{\Gamma(a^-)}\right]\,,
\label{coeff1}  \\
C\frac{(2\varepsilon)^{1-\tilde l}}{\Gamma(a)}\left(\varepsilon
  -\frac{1+a-b}{2-b}\right) &=&D \left[\frac{(2\varepsilon^+)^{1-\tilde
  l}}{\Gamma(a^+)} \left(\varepsilon^+ - \frac{1+ a^+ -b}{2 -b}\right)
  +\frac{e^{i\omega}(2\varepsilon^-)^{1-\tilde l}}{\Gamma(a^-)}
  \left(\varepsilon^- - \frac{1+ a^- -b}{2 -b}\right)\right]\,,
\label{coeff2} \\
C\frac{(2\varepsilon )^{\tilde l +1}}{\Gamma(1+a-b)}& =&D\left[
\frac{(2\varepsilon^+)^{\tilde l +1}}{\Gamma(1+a^+ -b)}
+\frac{e^{i\omega}(2\varepsilon^-)^{\tilde l +1}}{\Gamma(1+a^- -b)}\right]\,.
\label{coeff3}
\end{eqnarray}
Now we should mention one point regarding Eq. (\ref{coeff1}),
  Eq. (\ref{coeff2}) and Eq. (\ref{coeff3}) here. If we would  have written
  the Hamiltonian Eq. (\ref{rham}) near the origin like
\begin{eqnarray}
\mathcal H_{\tilde l}(r) = -\frac{1}{2}\frac{d^2}{dr^2} +\frac{\tilde l(\tilde
l+1)}{2r^{2}}\,,
\label{rham2}
\end{eqnarray}
neglecting the $\frac{1}{r}$ term compared to $\frac{\tilde l(\tilde
  l+1)}{2r^{2}}$ which is usually done in case of Hydrogen atom problem \cite
  {abro} to find out the behavior of the wave function near $r=0$, we would
  get the solution  of the form
\begin{eqnarray}
\chi(2\varepsilon r)\sim  A(2\varepsilon )^{-\tilde l} + B(2\varepsilon
)^{1+\tilde l}\,.
\label{waveorigin}
\end{eqnarray}
Keeping Eq (\ref{waveorigin}) in mind  we will just compare between
Eq. (\ref{coeff1}) and  Eq. (\ref{coeff3}), and the result is given by
\begin{eqnarray}
f(E)\equiv \frac{\Gamma(1+a-b)}{\Gamma(a)}\varepsilon^{-b+1} =\frac{\chi_1
    \cos(\omega/2-\frac{\pi \tilde l}{4}+\beta_1)}{\chi_2\cos(\omega/2
    +\frac{\pi \tilde l}{4}+\frac{\pi}{4} +\beta_2)},
\label{compare}
\end{eqnarray}
where $\Gamma(a^\pm) = \chi_1 2^{\frac{\tilde l}{2}} e^{\pm i\beta_1}$ and
$\Gamma(1+a^\pm-b)=\chi_2 2^{-\frac{\tilde l +1}{2}} e^{\pm i\beta_2}$.  We
can calculate the eigenvalue analytically when the right hand side is $0$ or
$\infty$. When $R.H.S = 0$, ie. when $\omega= \pi + \frac{\pi\tilde l
}{2}-2\beta_1 $, we get from left hand side
$\Gamma(a)=\Gamma(j+1+\frac{\delta_1+\delta_2}{2}-\frac{1}{\varepsilon}) =
\Gamma(-n)$, which implies
\begin{eqnarray}
E_n= - \frac{1}{2(n+1+j+\frac{\delta_1+\delta_2}{2})^2}\,; n=0,1,2,\cdots
\label{eigenvalue}
\end{eqnarray}
On the other hand when $R.H.S = \infty$, ie. when $\omega= \pi-\frac{\pi b}{4}
+\frac{\pi}{2}-2\beta_2 $, we get from left hand side
$\Gamma(1+a-b)=\Gamma(-j-1-\frac{\delta_1+\delta_2}{2}-\frac{1}{\varepsilon})
=\Gamma(-n)$, which implies\\
\begin{eqnarray}
E_n= - \frac{1}{2(n-1-j-\frac{\delta_1+\delta_2}{2})^2}\,; n=0,1,2,\cdots
\label{eigenvalue1}
\end{eqnarray}
For values other than $0$ and $\infty$ of the r.h.s of Eq.
(\ref{compare}) one can calculate the eigenvalues numerically like
Ref.~\cite{pisani1,pisani2,kumar}. But we are not going to plot it in our
work. Rather we make the following observations on our self-adjoint 
extensions of
generalized MIC-Kepler system.
\subsection{Case 1}
Consider the situation $\tilde l=j+ \frac{\delta_1+\delta_2}{2}= 0$.  This
correspond to the effective potential
\begin{eqnarray}
V_0= -\frac{1}{r}.
\end{eqnarray}
This happens when $j=\delta_1=\delta_2 =0$. $\delta_1$ and $\delta_2$ can be
made zero by choosing $c_1=c_2 =0$ and $j=0$ can be made  by choosing $s=0$
and considering lowest value of angular momentum, as evident from
Eq. (\ref{m}), Eq. (\ref{jvalue}) and Eq.  (\ref{angquantum}). But for
$s=c_1=c_2 =0$, generalized MIC-Kepler system reduces to Hydrogen atom
problem. Since $\tilde l=0$ lies inside the range of Eq. (\ref{msquareint}), we
can have 1-parameter family of self-adjoint extensions.  The above constraint
when substituted in
Eq. (\ref{eigenvalue}) gives the well known  Bohr level
\begin{eqnarray}
E_n= - \frac{1}{2(n+1)^2}\,; n=0,1,2,\cdots
\label{bhor}
\end{eqnarray}
which corresponds to the self-adjoint extension parameter  $\omega= \pi +
\frac{\pi\tilde l }{2}-2\beta_1(j=\delta_1=\delta_2=0)$. One point should be
mentioned here that the condition $j=\delta_1=\delta_2=0$  will not make Eq
(\ref{compare}) well behaved for negative integral values of $a$, because that
makes the l.h.s = $\frac{\infty}{\infty}$. So instead of putting  
$j=\delta_1=\delta_2=0$  directly we will consider the limit $b\to 2$ in Eq.
(\ref{compare}) to make it well behaved. Any deperture from  $\omega= \pi +
\frac{\pi\tilde l }{2}-2\beta_1(j=\delta_1=\delta_2=0)$
shifts the Bohr levels either to higher energy levels or lower energy levels
depending upon what values of $\omega$ are being considered. This shifts from
the usual bohr levels is called the Rydberg correction or the quantum defect
\cite{few} in atomic physics. The modified Bohr levels can be written in the
form 
\begin{eqnarray}
E_n= - \frac{1}{2(n+1 + \Delta_1)^2}\,; n=0,1,2,\cdots
\label{bhor1}
\end{eqnarray}
where $\Delta_1$, which plays the role of quantum defect,  depends upon the 
self-adjoint extension parameter $\omega$.

\subsection{Case 2}
Consider the situation $\tilde l=j+ \frac{\delta_1+\delta_2}{2}\neq 0$, but
$\tilde l< \frac{1}{2}$. This correspond to the effective potential
\begin{eqnarray}
V_{\tilde l}= -\frac{1}{r}+ \frac{\tilde l(\tilde l+1)}{r^2},~~~\tilde l<
\frac{1}{2}.
\end{eqnarray}
This happens when $j=0$ and $\delta_1+\delta_2 <1$. By appropriately choosing
$c_1$ and $c_2$, sum of two delta can be made less that one.  $j=0$ can be
made  by choosing $s=0$ and considering lowest value of angular momentum, as
evident from Eq. (\ref{m}), Eq.  (\ref{jvalue}) and
Eq. (\ref{angquantum}). But for $s=0$ and $c_1,c_2 \neq 0$, generalized
MIC-Kepler system reduces to Kepler-Coulomb problem, when $c_1\neq c_2$. For
$s=0$ and $c_1=c_2 \neq 0$, generalized MIC-Kepler system reduces to Hartmann
system. In these two cases also  we can have 1-parameter family of
self-adjoint extensions. In these two cases the spectrum for which 
the r.h.s of Eq. (\ref{compare}) is zero is given by 
\begin{eqnarray}
E_n= - \frac{1}{2(n+1 +\frac{\delta_1+\delta_2}{2})^2}\,; n=0,1,2,\cdots
\label{hart}
\end{eqnarray}
where $\omega= \pi +
\frac{\pi\tilde l }{2}-2\beta_1(j=0, c_1,c_2 \neq 0)$. For other values of
$\omega$ the spectrum should be modified and it can be written as 
\begin{eqnarray}
E_n= - \frac{1}{2(n+1 +\frac{\delta_1+\delta_2}{2}+\Delta_2)^2}\,; n=0,1,2,\cdots
\label{hart1}
\end{eqnarray}
where $\Delta_2$ is the quantum correction.
\subsection{Case 3}
Consider the situation $\tilde l=j+ \frac{\delta_1+\delta_2}{2}\neq 0$, but
$\tilde l> \frac{1}{2}$. This correspond to the effective potential
\begin{eqnarray}
V_{\tilde l}= -\frac{1}{r}+ \frac{\tilde l(\tilde l+1)}{r^2},~~~\tilde l>
\frac{1}{2}.
\end{eqnarray}
In this case since the deficiency space solutions are not square integrable,
deficiency indices become $<0,0>$. So there is no self-adjoint extensions. The
system is essentially self-adjoint. For charge-dyon case and MIC-Kepler case,
$s\ne0$; and  minimum value which $j$ can take is $j= |s|$.  
So in these cases the
corresponding radial Hamiltonian operator is essentially self-adjoint
everywhere. For essentially self-adjoint case, the unique spectrum is given by
\begin{eqnarray}
E_n= - \frac{1}{2(n+1+ j +\frac{\delta_1+\delta_2}{2})^2}\,; n=0,1,2,\cdots
\label{uniquespec}
\end{eqnarray}
%
%%%%%%%%%%%%%%%%%%%%%%%%%
\section{\small{\bf{Conclusion}}}\label{con}
In conclusion, self-adjointness of the quantum observables is an  important
issue. So when an operator is not self-adjoint in a domain one should search
for the possible self-adjoint extensions.  Because, it  not only  ensures 
the eigenvalue to be real but also through this we
can explore new spectrum which was not possible in usual analysis. It has been
successfully applied in different quantum mechanical models
\cite{few,biru1,biru2,reed,dunford,kumar}. In our present work we have 
considered
generalized MIC-Kepler system and we have found out a 1-parameter family of
self-adjoint extensions of the system for $\tilde l <\frac{1}{2}$. we have
shown that $\tilde l =0$ has a 1-parameter family of
self-adjoint extensions. $\tilde l =0$, corresponds to the orbital angular
momentum quantum no $l=0$ for the Hydrogen atom problem. For all other values
of the orbital angular momentum quantum no ($l\neq 0$), Hydrogen atom radial
Hamiltonian is essentially self-adjoint.

In case of Kepler-Coulomb system with $j=0$ and $\delta_1+\delta_2<1$
one gets again a  1-parameter family of self adjoint extensions. For all other values of
angular momentum ($j\neq 0$), the system is essentially self adjoint.

Hartmann system with $j=0$ and $\delta_1+\delta_2<1$  also has a
1-parameter family of self adjoint extensions. For all other values of angular
momentum ($j\neq 0$), the system is essentially self adjoint.

Since for Charge-dyon system and MIC-Kepler system,  minimum value of $j$ is
$\frac{1}{2}$, we don't have  self-adjoint extensions. System is essentially
self-adjoint in these cases.
%%%%%%%%%%%%%%%%%%%%%%%%%

%%%%%%%%%%%%%%%%%%%%%%%%%%%
\subsubsection*{Acknowledgments}
%%%%%%%%%%%%%%%%%%%%%%%%%%%
We thank  B. Basu-Mallick, Kumar S. Gupta and Palash B. Pal for comments on the
manuscript and helpful discussions.
%%%%%%%%%%%%%%%%%%%%%%%%%%%

\end{document}